# Recent results from MicroBooNE

**Holly B. Parkinson**[a,*] **for the MicroBooNE Collaboration**

[a]*University of Edinburgh,*
 *Edinburgh EH9 3FD, United Kingdom*

 *E-mail:* h.b.parkinson@sms.ed.ac.uk

Modelling and reconstructing neutrino-nucleus scattering is difficult, but it is crucial to do it precisely to enable next-generation oscillation measurements. Liquid argon time projection chambers (LArTPCs), such as MicroBooNE, can be the tool for this job as they are excellent precision neutrino detectors with their ability to produce detailed three-dimensional interaction images and precise energy and spatial resolution. MicroBooNE currently possesses the world's largest neutrino-argon scattering data set collected over five years using the BNB and NuMI neutrino beams at Fermilab. The experiment has performed measurements over a broad range of physics topics including neutrino argon cross sections, searches for BSM physics, and investigations of the MiniBooNE LEE excess. Many of these measurements are essential for improving the modelling of nuclear effects for both MicroBooNE and future LArTPC neutrino experiments, such as DUNE. This talk will give an overview of recent MicroBooNE results, the analysis techniques that enable them, and prospects for future measurements.

*Speaker

## 1. Introduction

The Micro Booster Neutrino Experiment (MicroBooNE) is a Liquid Argon Time Projection Chamber (LArTPC) neutrino detector. MicroBooNE is located at Fermilab in Batavia, Illinois, and is part of Fermilab's Short Baseline Neutrino (SBN) Program. The experiment operated from 2015 to 2021 and has collected a large, well-understood dataset of neutrino-argon interactions; the detector is currently in its decommissioning phase.

MicroBooNE is positioned uniquely such that it is able to receive neutrinos from two beams: it is on axis with the Booster Neutrino Beam (BNB) at 470 m from the target, and approximately 680 m and 8 degrees off axis from the Neutrinos at the Main Injector beam (NuMI). The main goals of MicroBooNE are to investigate the MiniBooNE low energy excess, measure low-energy neutrino cross sections, and provide input for the building and operation of future LArTPC experiments, such as the Deep Underground Neutrino Experiment (DUNE).

The MicroBooNE Collaboration consists of over 180 collaborators from 40 institutions, and, at the time of writing, has published 70 papers.

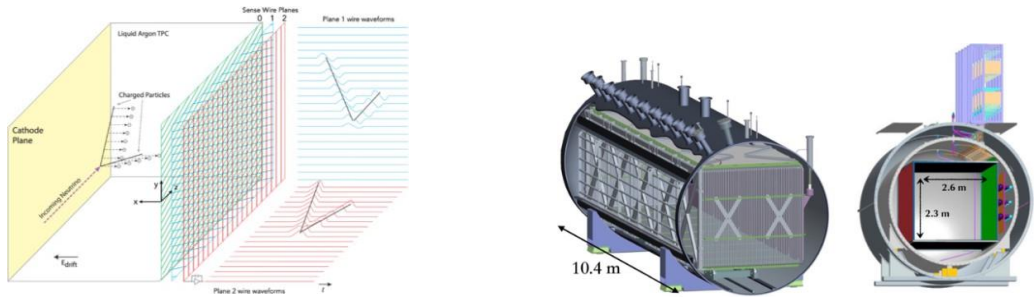

**Figure 1:** Left: Diagram showing the operating principle of a LArTPC, with electric field direction, coordinate system and drifting charged particles labelled. The green and blue wire planes are the 'induction planes', and the red wire plane is the 'collection plane'; Right: Schematic view of the MicroBooNE detector.

## 2. MicroBooNE as a LArTPC

In a LArTPC, a neutrino interaction with argon can produce charged particles, which can go on to ionize the argon and produce ionisation electrons and scintillation light. The charge is drifted with an electric field and collected on wire planes to precisely reconstruct interaction positions and perform calorimetry; the light is collected by photodetectors and is used to assign a time to the interaction and reject non-beam background. MicroBooNE has an 85 tonne active volume, 3 planes of wires (vertical, +60°, -60°) with 3 mm spacing for charge collection, and 32 PMTs to detect scintillation photons. The detector is capable of mm-level spatial resolution, and can produce 3D interaction images. It has a fully active tracking calorimeter to achieve precise energy resolution, excellent particle identification (including the ability to distinguish electrons from photons based on shower energy deposition) [1], and a Cosmic Ray Tagger (CRT) installed around its cryostat to improve cosmic background rejection [2].

MicroBooNE has a history of developing analysis tools for LArTPC physics, including the achievement of a neutrino interaction time resolution of O(1 ns), which will improve cosmic ray rejection for neutrino analyses and aid in searches for beyond standard model physics [3]. Reconstruction advancements made by MicroBooNE have allowed for MeV-scale reconstruction, aiding in low-energy calorimetry, and particle identification, such as the separation of muon signals from pion signals [4]. MicroBooNE has also demonstrated the ability to identify neutrons in neutrino interactions, a technique that can be applied to any LArTPC [5].

These analysis techniques are important to aid the construction of future large-scale LArTPC detectors such as those under construction for DUNE. Post-operation detector R&D studies of MicroBooNE are currently ongoing to investigate the state of the detector after its years of use.

## 3. Physics at MicroBooNE

### 3.1 The MiniBooNE Low Energy Excess

The Mini Booster Neutrino Experiment observed a $4.8\sigma$ excess of electron neutrino-like events at low energies compared to prediction [6]; a $3.8\sigma$ excess of electron antineutrino events had been seen earlier by the Liquid Scintillator Neutrino Detector (LSND). MiniBooNE and LSND were both Cherenkov detectors and could not distinguish between electrons and photons, but a LArTPC can, giving MicroBooNE the ability to investigate the nature of the excess as one of its primary physics goals.

In 2022, MicroBooNE published results disfavouring an electron-like explanation for the LEE using 3 charged current electron neutrino searches with four event classes (1e0p, 1e1p, 1eNp, 1eX), and an electron excess was rejected at $> 97\%$ CL [7]. MicroBooNE's new 2024 analysis uses the full MicroBooNE dataset ($1.11 \times 10^{21}$ POT), and uses two exclusive samples of electron neutrinos without visible pions, one with visible protons and one without any visible protons (1e0p0$\pi$ and 1eNp0$\pi$). The analysis includes the CRT in event selections, uses a new MiniBooNE LEE empirical model, and represents the LEE as a function of shower energy and angle. The electron excess was rejected at $> 99\%$ CL in all kinematic variables.

Beyond Standard Model explanations for the LEE have also been explored by MicroBooNE, such as 3+1 sterile neutrino oscillations. Using data from $6.37 \times 10^{20}$ POT, no evidence of 3+1 sterile neutrino oscillations was observed; MicroBooNE has reported prospects of combining BNB and NuMI data to significantly improve sensitivity. The new analysis will be sensitive to more LSND parameter space [8]. Dark sector neutrinos decaying into overlapping/asymmetric $e^+ e^-$ pairs could lead to a signature mimicking the excess, and MicroBooNE has presented substantial improvements in efficiency in exploring new dark sector parameter space relative to first-generation analyses, aiming to confirm or reject dark sector models as an explanation for the MiniBooNE LEE when a larger data sample is made available.

### 3.2 Beyond Standard Model (BSM) Physics

In a neutrino beamline, kaons and pions are produced to decay into neutrinos, but these could potentially decay into other species, providing avenues for BSM searches. MicroBooNE has world leading limits in searches for new particles in O(10 MeV) – 300 MeV range under several phenomenological models.

MicroBooNE has reported the strongest limit to date on the mixing angle $\theta$ for a new scalar particle $S$ mixing with the Higgs field in the mass range 110 MeV < $m_S$ < 155 MeV, searching for Higgs-portal scalar particles decaying through the $S \rightarrow e^+ + e^-$ channel. The scalar particles are produced from kaons decaying both in flight and at rest. A limit was set of $\theta < 2.48 \times 10^{-4}$ ($\theta < 1.60 \times 10^{-4}$) at $m_S$ = 125 MeV ($m_S$ = 150 MeV) at the 95% confidence level.

A search for heavy neutral leptons (HNLs) (produced from $K^+$ decays) decaying into $\mu^\pm \pi^\mp$ pairs at MicroBooNE has resulted in an upper limit for the mixing parameter $|U_{\mu 4}|^2$ of $|U_{\mu 4}|^2$ = 12.9 x $10^{-8}$ (0.92 x $10^{-8}$) for $m_{HNL}$ = 246 MeV (385 MeV). This is an order of magnitude improvement on previous MicroBooNE results, and achieved a similar sensitivity to the NA62 experiment [9].

### 3.3 Neutrino-Argon Cross Sections

MicroBooNE possesses a large, well-understood neutrino-argon interaction dataset after 5 years of data taking, and has published over 20 $\nu$-Ar cross sections using various channels.

$\pi^0$ are an important background in $\nu_e$ searches, as a $\pi^0$ interaction produces two showers, but if one is missed (likely due to one shower leaving the detector, or the two showers being on top of each other), it may resemble a $\nu_e$ interaction. In 2024, MicroBooNE has presented the first double-differential cross section measurement in $\cos(\theta_{\pi^0})$ and $P_{\pi^0}$ of neutral-current neutral pion (NC$\pi^0$) production in $\nu$-Ar scattering, along with single-differential measurements in $P_{\pi^0}$ using final states with or without protons. 4971 NC$\pi^0$ events were selected for this analysis [10]. MicroBooNE has also presented measurements of differential cross sections of neutral pion production in charged-current muon neutrino interactions (CC$\pi^0$). Differential cross sections in muon momentum, neutrino-muon scattering angle, muon-pion opening angle, pion momentum, and pion detection angle are measured, using $6.86 \times 10^{20}$ POT of data. The measurements have been compared to several generator predictions and a good agreement is reported, except for an overprediction at muon forward angles, giving scope for the generators to be improved [11].

MicroBooNE has used generalized kinematic imbalance (GKI) variables derived from longitudinal components along the beam direction to achieve improved sensitivity to nuclear effects, which are important for reducing uncertainty in precise modelling of neutrino-nucleus interactions. The first flux-integrated single and double-differential cross section measurements in these variables have also been measured, using $\nu_\mu$-Ar CC1p0$\pi$ interactions, and a clear interaction model separation has been shown. These measurements could be used to tune neutrino-nucleus interaction models, especially modelling of final state interactions [12].

Due to being off-axis, NuMI provides MicroBooNE a higher flux of $\nu_e$ and $\bar{\nu}_e$, which is useful for measuring electron-type neutrino cross sections (BNB has smaller $\nu_e/\bar{\nu}_e$ content, but exclusive measurements using BNB are still possible). Inclusive measurements of $\nu_e + \bar{\nu}_e$ cross sections have been performed, and exclusive $\nu_e$ and $\bar{\nu}_e$ measurements are in progress [13] [14]. Measurements of electron-type neutrino cross sections using the full MicroBooNE dataset are also in progress.

### 3.4 Neutron identification

MicroBooNE's most recent paper (at the time of writing) demonstrates tagging of neutrons produced in neutrino interactions. The neutrons were found using observations of secondary protons produced in their interaction with argon some distance from the neutrino vertex, and the study aims to overcome missing energy issues caused by undetected neutrons in LArTPCs. This technique could allow measurement of neutron production from neutrino interactions, and could provide statistical separation between neutrinos and antineutrinos, which are otherwise difficult to separate in LArTPCs. This neutron identification technique can be applied to any LArTPC. The sample used, a small subset of MicroBooNE data, had a purity of 60%, which yielded an integrated efficiency of 8.4% for neutrons that produce a detectable proton, but there are prospects for efficiency improvement. It is noted that in a larger detector, such as DUNE, this method will yield a higher efficiency.

## 4. Conclusion

MicroBooNE is a LArTPC neutrino detector that operated at Fermilab from 2015 - 2021, and has collected a large, well-understood neutrino-argon interaction dataset. The collaboration is very active, with recent results in several areas of physics; further analyses aim to utilise the full dataset, incorporate NuMI and BNB data together, and implement an updated NuMI flux. The detector is currently in a decommissioning R&D phase with results to come soon. The advancements and measurements made by MicroBooNE have contributed to a greater understanding of neutrino interactions and to the construction of future LArTPC detectors.

## Acknowledgements

This document was prepared by the MicroBooNE collaboration using the resources of the Fermi National Accelerator Laboratory (Fermilab), a U.S. Department of Energy, Office of Science, HEP User Facility. Fermilab is managed by Fermi Research Alliance, LLC (FRA), acting under Contract No. DE-AC02-07CH11359. MicroBooNE is supported by the following: the U.S. Department of Energy, Office of Science, Offices of High Energy Physics and Nuclear Physics; the U.S. National Science Foundation; the Swiss National Science Foundation; the Science and Technology Facilities Council (STFC), part of the United Kingdom Research and Innovation; the Royal Society (United Kingdom); the UK Research and Innovation (UKRI) Future Leaders Fellowship; and the NSF AI Institute for Artificial Intelligence and Fundamental Interactions. Additional support for the laser calibration system and cosmic ray tagger was provided by the Albert Einstein Center for Fundamental Physics, Bern, Switzerland. We also acknowledge the contributions of technical and scientific staff to the design, construction, and operation of the MicroBooNE detector as well as the contributions of past collaborators to the development of MicroBooNE analyses, without whom this work would not have been possible. For the purpose of open access, the authors have applied a Creative Commons Attribution (CC BY) public copyright license to any Author Accepted Manuscript version arising from this submission.